\documentclass{iopart}
\usepackage{graphicx,amsbsy,bbm,amsmath,iopams}
\newcommand{\X}{{\rm X}} \newcommand{\A}{{\rm A}}
\newcommand{\B}{{\rm B}} \newcommand{\C}{{\rm C}}
\newcommand{\D}{{\rm D}} \newcommand{\bfX}{\mathbf X} 
\newcommand{\bX}{\overline{\rm X}} \newcommand{\Y}{{\rm Y}}
\newcommand{\bY}{\overline{\rm Y}}
\newcommand{\tru}{{\rm Tr}_1} \newcommand{\trd}{{\rm Tr}_2}
\newcommand{\trud}{{\rm Tr}_{1,2}}
\newcommand{\vp}{{\varrho_p}} \newcommand{\iid}{\mathbb{I}}
\newcommand{\rmSU}{{\rm SU}}
\newcommand{\si}{{\sigma_0}} \newcommand{\su}{{\sigma_1}}
\newcommand{\sd}{{\sigma_2}} \newcommand{\st}{{\sigma_3}}
\newcommand{\refeq}[1]{Eq.~(\ref{#1})} \newcommand{\ket}[1]{\vert #1 \rangle}
\newcommand{\bra}[1]{\langle #1 \vert} \newcommand{\avg}[1]{\langle#1\rangle}
  
\newtheorem{teo}{Theorem}
\begin{document}
\title{Cloning of observables}
\author{Alessandro Ferraro$^{\dag,\circ}$, Matteo Galbiati$^\ddag$, Matteo G. A.
Paris$^\dag$}
\address{$^\dag$ Dipartimento di Fisica dell'Universit\`a di Milano, Italia.}
\address{$^\ddag$ STMicroelectronics, I-20041 Agrate Brianza (MI), Italia.}
\address{$^\circ$ Institut de Ci\`encies Fot\`oniques, E-08860 Castelldefels, 
(Barcelona), Spain} 
\date{\today}
\begin{abstract}
We introduce the concept of cloning for {\em classes of
observables} and classify cloning machines for qubit systems
according to the number of parameters needed to describe the
class under investigation. A no-cloning theorem for observables
is derived and the connections between cloning of observables and
joint measurements of noncommuting observables are elucidated.
Relationships with cloning of states and non-demolition
measurements are also analyzed. 
\end{abstract}
\section{Introduction}\label{s:intro}
Information may be effectively manipulated and transmitted
by encoding symbols into quantum states. However, besides 
several advantages, the quantum nature of the transmitted
signals entails some drawback, the most relevant owing to 
the so-called {\em no-cloning} theorem: Quantum 
information encoded in a set of nonorthogonal states cannot 
be copied \cite{nc1,nc2,nc3}. 
In order to overcome this limitation an orthogonal coding may 
be devised, which however requires the additional control of 
the quantum channel since, in general, propagation 
degrades orthogonality of any set of input quantum signals.
\par
A different scenario arises by addressing transmission of
information encoded in the statistics of a set of observables,
independently on the quantum state at the input. In a network of
this sort there is no need of a precise control of the coding
stage whereas, on the other hand, each gate should be {\em
transparent}, {\em i.e.} should preserve the statistics of the
transmitted observables. In this paper, we address the problem of
copying information that has been encoded in the statistics of a set
of observables. For this purpose we introduce the concept of
cloning machine for classes of observables and analyze in details
the constraint imposed by quantum mechanics to this kind of
devices.  Two forms of a no-cloning theorem for observables will
be derived, and the connections with cloning of states and joint
measurements will be discussed.
\par
In the following we assume that information is encoded in the
statistics of a set of qubit observables. A cloning machine for the
given set is a device in which a signal qubit carrying the
information interacts with a  probe qubit via a given unitary
with the aim of reproducing the statistics of each observable on {\em
both} the qubits at the output. The chance of achieving this
task, besides the choice of a suitable interaction, depends on the
class of observables under investigation. In this paper we
provide a full classification of cloning machines, based on the
number of parameters needed to specify the class.
\par
The paper is structured as follows: In Section \ref{s:CloM} we
introduce the concept of cloning machine for a class of
observables and illustrate some basic properties. Then, the
possibility of realizing a cloning machines for a given class is
analyzed, depending on the number of parameters individuating the
class.  The results are summarized in two forms of a no-cloning
theorem for observables. In Section \ref{s:CloJM} we analyze the
connections between cloning of observables and joint measurements
of noncommuting observables. Section \ref{s:outro} closes the
paper with some concluding remarks.
\section{Cloning of observables}
\label{s:CloM}
\begin{figure}[h]
\begin{center}
\includegraphics[width=0.7\textwidth]{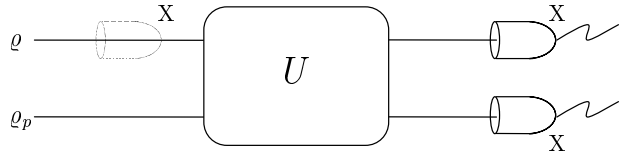}
\caption{Schematic diagram of a cloning machine for observables 
$(U,\vp,\bfX)$: a signal qubit 
prepared in the unknown state $\varrho$ interacts, via 
a given unitary $U$, with a probe qubit prepared in the 
known  state $\vp$. The class of observables
$\bfX$ is cloned if a measurement of any $\X\in\bfX$ on
either the two qubits at the output gives the same statistics as
it was measured on the input signal qubit, {\em independently} on
the initial qubit preparation $\varrho$.}
\label{f:machine}
\end{center}
\end{figure}

We consider a device in which a signal qubit (say, qubit "1") is
prepared in the (unknown) state $\varrho$ and then interacts, via 
a given unitary $U$, with a probe qubit ("2") prepared in the 
known  state $\vp$. For a given class of qubit observables $\bfX
\equiv \left\{ \X(j)\right\}_{j\in{\cal J}}$ where ${\cal J}$ is
a subset of the real axis and $\X(j) \in {\cal L}[{\mathbb
C}^2]$, we introduce the concept of cloning as follows.  A
cloning machine for the class of observables $\bfX$ is a triple
$(U,\vp,\bfX)$ such that 
$$ \bX_1 = \bX \quad \bX_2=\bX \qquad \forall  \varrho  
\quad \forall\: \X \in \bfX\:,$$ 
where $\bX \equiv \hbox{Tr}_1 \left[ \varrho\: \X\right]$
is the mean value of the observable $X$ at the input
and 
\begin{eqnarray}
\bX_1 \equiv \hbox{Tr}_{12}\left[ R\: 
\X \otimes I \right] \quad \bX_2 \equiv
\hbox{Tr}_{12}\left[ R\: I \otimes \X\right]
\end{eqnarray}
are the mean values of the same observable on the two output
qubits (see Fig. \ref{f:machine}).
The density matrix $R= U\: \varrho \otimes \vp \:U^\dag$
describes the (generally entangled) state of the two qubits after the
interaction, whereas $I$ denotes the identity operator.  
\par
The above definition identifies the cloning of an observable with
the cloning of its mean value. This is justified by the fact that
for any single-qubit observable $\X$ the cloning of the mean
value is equivalent to the cloning of the whole statistics. In
fact, any $\X \in {\cal L}[{\mathbb C}^2]$ has at most two
distinct eigenvalues $\{\lambda_0, \lambda_1\}$, occurring with
probability ${p_0,p_1}$. For a degenerate eigenvalue the
statement is trivial. For two distinct eigenvalues we have $\bX =
\lambda_1 p_1 + \lambda_0 p_0$ which, together with the
normalization condition $1=p_0+p_1$, proves the statement.  In
other words, we say that the class of observables $\bfX$ has been
cloned if a measurement of any $\X\in\bfX$ on either the two
qubits at the output gives the same statistics as it was measured
on the input signal qubit, {\em independently} on the initial
qubit preparation.  
\par
A remark about this choice is in order. In fact, in view of the 
duality among states and operators on an Hilbert space, one may 
argue that a proper figure of merit to assess a cloning machine 
for observables would be a fidelity--like one. This is certainly 
true for the $d$-dimensional case, $d>2$, while for qubit 
systems a proper assessment can be also made in term of mean--value 
duplication, which subsumes all the information carried by the signal.
\par
Our goal is now to classify cloning machines according to the number
of parameters needed to fully specify the class of observables 
under investigation. Before beginning our analysis let us illustrate
a basic property of cloning machines, which follows 
from the definition, and which will be extensively used throughout 
the paper. \par 
Given a cloning machine $\left(U,\vp,\bfX\right)$, then
$\left(V,\vp, {\mathbf Y} \right)$
is a cloning machine too, where  
$V=(W^\dag \otimes W^\dag) \:U \:(W\otimes I)$ 
and the class ${\mathbf Y}=W^\dag \bfX W$ is formed by 
the observables $Y(j)=W^\dag X(j)\,W$, $j\in{\cal J}$.
The transformation $W$ may be a generic unitary.
We will refer to this property to as {\em unitary covariance}
of cloning machine. The proof proceeds as follows.
By definition $\bY(j)=\tru[\varrho\: W^\dag X(j) W] 
=\tru[W\,\varrho\,W^\dag\,\X(j)]$. Then, since 
$\left(U,\vp,\bfX \right)$ is a cloning machine, we have 
\begin{align}
\bY(j) &=\trud[U\,(W\varrho\, W^\dag \otimes \vp)\,U^\dag\,(\X(j)
\otimes \iid)] \nonumber \\ &=\trud[U\,(W\otimes \iid)(\varrho
\otimes \vp)(W^\dag \otimes \iid)\,U^\dag\: (W\otimes W)(\Y(j)
\otimes \iid)(W^\dag\otimes W^\dag)] \nonumber \\
&=\trud[V\,(\varrho \otimes \vp)\,V^\dag\,(\Y(j) \otimes
\iid)]=\bY_1(j)\;.  
\end{align}
The same argument holds for $\bY_2(j)$.
\par
Another result which will be used in the following is the 
parameterization of a two-qubit transformation, 
which corresponds to a $\rmSU (4)$ matrix, 
obtained by separating its local and entangling parts. A 
generic two-qubit gate $\rmSU (4)$ matrix may be factorized 
as follows 
\cite{car}:
\begin{align}\label{SU4}
  U=L_2\,U_{\rm E}\,L_1=L_2\,
\exp\Big[\frac{i}{2}\sum_{j=1}^{3}\theta_j\sigma_j\otimes\sigma_j\Big]
\,L_1
\end{align}
where $\theta_j \in \mathbb{R}$ and the $\sigma_j$'s are the Pauli's
matrices. The local transformations $L_1$ and $L_2$ belongs to the 
$\rmSU (2)\otimes \rmSU (2)$ group, whereas $U_{\rm E}$ accounts for 
the entangling part of the transformation $U$. 
In our context, decomposition (\ref{SU4}), together with unitary
covariance of cloning machines, allows to ignore the local 
transformations $L_1$, which corresponds 
to a different state preparation of signal and probe qubits at 
the input. On the other hand, as we will see in the following, the degree
of freedom offered by the local transformations $L_2$ will be exploited
to design suitable cloning machines for noncommuting observables.
\subsection{One--parameter classes of observables}
Let us begin our analysis with a class constituted by only one
observable $\A$. In this case a cloning machine $(U,\vp,\A)$
corresponds to a quantum non-demolition measurement of $\A$
itself, {\em i.e.} a measurement which introduces no back-action
on the measured observable, thus allowing for repeated
measurements \cite{cav}.  As an example, if $\A=\st$ then $(U_C,
|0\rangle\langle 0|, \st)$ is a cloning machine \cite{ral}, $U_C$
being the unitary performing the C$_{not}$ gate.  The proof is
straightforward since 
$$\trd[(\iid\otimes |0\rangle\langle 0|)\,U_C^\dag\,(\st\otimes\iid)\,U_C]=
\trd[(\iid\otimes |0\rangle\langle
0|)\,U_C^\dag\,(\iid\otimes\st)\,U_C]=\st\:. $$
The next step is to consider a generic one--parameter class of 
observables. At first we notice that $(U_C, |0\rangle\langle 0|, \bfX_3)$ 
is a cloning machine for the class $\bfX_3 = \{ x_3 \st\}_{x_3\in {\mathbb
R}}$. Using this result, and denoting by $\sigma_0$ the identity matrix, 
we find a cloning machine for the class $\bfX_\A \equiv \left\{x\A 
\right\}_{x\in {\mathbb R}}$, where $\A=\sum_{k=0}^3 a_k\sigma_k$
is a generic observable. 
Explicitly, the triple 
\begin{eqnarray}
(U_\A,\ket{0}\bra{0},\bfX_\A) \qquad
U_\A=(W_\A^\dag \otimes W_\A^\dag) \:U_C \:(W_\A\otimes \iid)
\label{1pclass}\;
\end{eqnarray}
is a cloning machine, with the single-qubit unitary transformation 
$W_\A$ is given by $W_\A=\exp(i\phi\sum_{j=1}^{2}n_j\sigma_j)$, 
with $n_2^2=1-n_1^2$, $$n_1=\frac{a_2}{\sqrt{a_1^2+a_2^2}}\:,$$ 
and
$$\phi=\arccos \frac{a_3}{\sqrt{a_1^2+a_2^2+a_3^2}}\:.$$
In order to prove the statement one first notices that any observable 
of the form $\A' =\sum_{j=1}^{3} a _j \sigma_j$ can be obtained from 
$x_3\st$ by the unitary transformation $\A'=W^\dag x_3 \st W$
with $x_3=\sqrt{a_1^2+a_2^2+a_3^2}$ and $W=W_{\A}$.  The statement then 
follows from unitary covariance, since $\left(U_C,|0\rangle\langle 0|,
\bfX_3\right)$ is a cloning machine, whereas the identity matrix is 
trivially cloned.
\subsection{The set of all qubit observables}
Let us now consider the $n$--parameter class
$\bfX_g=\{x_1\,\X(1)+\dots+x_n\,\X(n)\}_{x_1,\dots,x_n\in{\mathbb R}}$.
Recall that the aim of a cloning machine for observables is to copy the
expectation value of a generic linear combination $\X$ of the $n$ 
observables $X(j)$ {\em i.e.} $\X=x_1\,\X(1)+\dots+x_n\,\X(n)$.
By decomposing each observable $\X(j)$ on the Pauli matrices basis 
$\X(j)=a_{j,0}\si+a_{j,1}\su+a_{j,2}\sd+
a_{j,3}\st$ and reordering one obtains:
  $\X=y_0\si+y_1\su+y_2\sd+y_3\st$, 
  where $y_k= \sum_{j=1}^n x_j\, a_{j,k}$. From the expression above we
see that at most a four--parameter class may be of interest, being any
other class of observables embodied in that. That being said, 
the following
operator counterpart of the usual no--cloning theorem for states can
be formulated:
\begin{teo}\label{noclo1}{\em\bf (No--cloning of observables I )}
  A cloning machine $\left(U,\vp,\bfX_g\right)$ where $\bfX_g$ is a generic
    $n$--parameter class of qubit observables does not exist.
\par\noindent Proof. {\rm Let us reduce to a four--parameter class as 
above. Then the request $\bX=\bX_1=\bX_2$ implies
$$\hbox{Tr}_{1}\left[ \varrho \: \sigma_k \right] =
\hbox{Tr}_{12}\left[ R\: \sigma_k \otimes I \right] =
\hbox{Tr}_{12}\left[ R\: I \otimes \sigma_k \right]\quad \forall k\:,$$ 
which, in turn, violates the no-cloning theorem for quantum states, since
expresses the equality of the input Bloch vector with that of the two
partial traces at the output. $\square$}
\end{teo}
The results of theorem \ref{noclo1} and the fact the the triple
in Eq. (\ref{1pclass}) is a cloning machine 
permit a comparison among cloning machines for observables and 
for states. Let us write the generic input signal as $\varrho = 
\frac12 (\sigma_0 + {\mathbf s} \cdot{\boldsymbol \sigma})$, where
${\boldsymbol \sigma}=(\sigma_1,\sigma_2,\sigma_3)$ and ${\boldsymbol
  s}=(s_1,s_2,s_3)$ is the Bloch vector, and consider the  cloning 
of the single--parameter class $\bfX_3$ via $(U_C,\ket{0}\bra{0},\bfX_3)$. 
The action of this cloning machine is the perfect copying of the
component $s_3$ of the Bloch vector $\boldsymbol s$, whereas the
values of $s_1$ and $s_2$ are completely disregarded. The same
situation occurs with any single--parameter class: a single
component of the generalized Bloch vector is copied, upon
describing the qubit in a suitable basis.  On the other hand,
when the whole class of qubit observables is considered, this
requirement should be imposed to all the components and cannot be
satisfied. Only approximate cloning is allowed for the entire
state of a generic qubit, with the whole Bloch vector being
shrunk by the same factor with respect to the initial Bloch
vector $\boldsymbol s$ \cite{buz,gis,wer,mac}. In the following
we analyze intermediate situations between the above two extrema
{\em i.e.} perfect copying (one-parameter class) and no-copying
(the set of all qubit observables).
\subsection{Two--parameter classes of observables}
\begin{figure}[h]
\begin{center}
\includegraphics[width=0.9\textwidth]{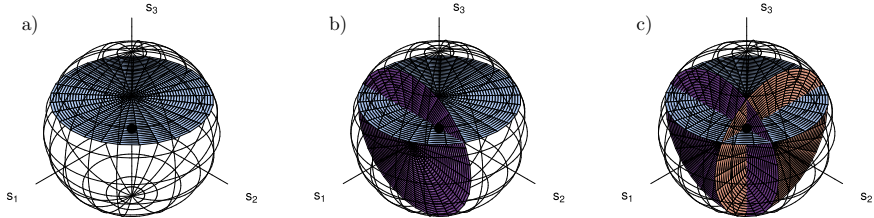}
\caption{Pictorial representation of the action of  
cloning machines for observables at the level of the Bloch 
sphere. 
In panel a): any cloning machine for the class $\bfX_3$
should preserve the third component of the Bloch vector, 
while modifying arbitrarily the other two. As a consequence, 
given a signal qubit at the input (denoted by a black circles) 
the Bloch vector of the two output qubits may lie on any point of
a plane. 
In panel b) we show the action of hypothetical
cloning machines for the class $\{x_1\su+x_3\st\}$, which 
would send the input qubit denoted by the black circle 
to output qubits lying on the intersection
of two planes: the meaning of Theorem \ref{noclo2} 
is that of preventing 
the existence of any such cloning machine. 
Finally, the no--cloning theorem for states does not 
allow the existence of any cloning machine for the class
$\{x_1\su+x_2\sd+x_3\st\}$, which for the input qubit denoted
by the black circle would impose two output states with
the same Bloch vector [corresponding to the intersection 
of the three surfaces as shown in panel c)].} \label{f:spheres} 
\end{center}
\end{figure}

In order to introduce cloning machines for 
two--parameter classes of observables, let us consider two specific 
classes: $\bfX_{\rm c_1}=\{x_0\si+x_3\st\}_{x_0,x_3\in\mathbb R}$ and
$\bfX_{\rm nc}=\{x_1\su+x_2\sd\}_{x_1,x_2\in\mathbb R}$. The first
class is constituted by commuting observables, hence one expects no 
quantum constraints on cloning them. This is indeed the case, and
an explicit representative of a cloning machine is given by
$(U_C,\ket{0}\bra{0},\bfX_{\rm c_1})$. The statement follows by
reminding that $(U_C,\ket{0}\bra{0},\bfX_3)$ is a cloning machine and
by noticing that the identity matrix $\si$ is trivially cloned by any
cloning machine. As already noticed such a cloning machine 
copies the component $s_3$ of the input signal Bloch vector $\boldsymbol s$.
On the other hand, consider the class $\bfX_{\rm nc}$, constituted by
noncommuting observables. If a cloning machine $(U,\vp,\bfX_{\rm nc})$
existed, then the mean values as well as the statistics of any
observable belonging to $\bfX_{\rm nc}$ would be cloned at its output. 
As a consequence, one would jointly
measure any two non-commuting observables belonging to $\bfX_{\rm nc}$
({\em e.g.}, $\su$ on the output signal and $\sd$ on the output probe)
without any added noise, thus violating the bounds imposed by quantum
mechanics \cite{ake,ago,y82}. 
Generalizing this argument to any two--parameter class of
noncommuting observables ({\em i.e.}, to any class
$\bfX_{\rm gnc}=\{c\,\C+d\,\D\}_{c,d\in{\mathbb R}}$, with $\C,\,\D$ generic
non-commuting observables), we then conclude with the following
stronger version of Theorem \ref{noclo1}:
\begin{teo}\label{noclo2}{\em\bf (No--cloning of observables II )}
  A cloning machine for a generic two--parameter class of noncommuting
  observables does not exist.
\end{teo}
The state--cloning counterpart of the Theorem above can be obtained by
considering the class of observables $\bfX_{\rm nc}$: If a cloning
machine $(U,\vp,\bfX_{\rm nc})$ existed, then the components $s_1$ and
$s_2$ of Bloch vector $\boldsymbol s$ would be cloned for any input
signal. The same situation occurs in the case of a two--parameter
class generated by any pair of Pauli operators. In other words, it is
not possible to simultaneously copy a pair of components of the Bloch
vector of a generic state, even completely disregarding the third one
\cite{kie,app}. In order to clarify the relationship between cloning 
of states and cloning of observables a pictorial view of the action 
of cloning machines for observables at the level of states is given 
in Fig. \ref{f:spheres}. 
\par
In Fig. \ref{f:spheres}a we show the action of cloning machines for 
the class $\bfX_3$: any cloning machine for this class
should preserve the third component of the Bloch vector, 
while modifying arbitrarily the other two. As a consequence, 
given a signal qubit at the input (denoted by a black circles) 
the Bloch vector of the two output qubits may lie on any point of
a plane of fixed latitude. 
In Fig. \ref{f:spheres}b we show the action of hypothetical
cloning machines for the class $\{x_1\su+x_3\st\}$: starting
from the input qubit denoted by the black circle the output
qubits would correspond to Bloch vectors having the same
first and third components, {\em i.e.} lying on the intersection
of two planes similar to that of Fig. \ref{f:spheres}a.
The meaning of Theorem \ref{noclo2} is that of preventing 
the existence of any such cloning machine. 
Finally, the no--cloning theorem for states does not 
allow the existence of any cloning machine for the class
$\{x_1\su+x_2\sd+x_3\st\}$, which for the input qubit denoted
by the black circle would impose two output states with
the same Bloch vector (corresponding to the intersection 
of the three surfaces).
\par
By theorem \ref{noclo2}, it is also clear that a three--parameter 
class of observables does not warrant further attention. 
In fact, a cloning machine for a three--parameter
class of noncommuting observables does not exist, whereas a
three--parameter class of commuting observables reduces to that
of a two--parameter class (of commuting observables). 
\par
\subsection{Class of commuting observables}
Concerning commuting observables, it
turns out that the cloning machine in Eq. (\ref{1pclass})
for a generic single--parameter class of observables also provides a 
cloning machine for a generic two--parameter class of commuting observables. 
In order to prove this statement let us first recall the
relationship between two generic commuting observables.
Given an observable $\A=\sum_{k=0}^3\,a_k\,\sigma_k$, then a generic
observable $\B$ commuting with $\A$ is given by
\begin{eqnarray}
\B=\sum_{k=0}^3\,b_k\,\sigma_k \qquad b_1=a_1 b_3/a_3\:\: b_2=a_2 b_3/a_3
\label{ABcomm}\;,
\end{eqnarray}
whereas $b_0$ and $b_3$ are free parameters \cite{nt1}.
Considering now two generic commuting observables $\A$ and $\B$, 
one has that a cloning machine for the class $\bfX_{\rm
    c}=\{a\,\A+b\,\B\}_{a,b\in\mathbb R}$ with $[\A,\B]=0$ is given by
  $(U_\A,\ket{0}\bra{0},\bfX_{\rm c})$, where $U_\A$ is given in
  (\ref{1pclass}).
The proof starts from the decomposition of $\A$ in the Pauli basis, 
namely $\A=\sum_{k=0}^3a_k\,\sigma_k$, and by defining $\A_{\rm r}=
a_0\si+a_3\st$. From (\ref{ABcomm}) one has that an observable $\B_{\rm r}$
  commuting with $\A_{\rm r}$ must be of the form $\B_{\rm
    r}=b_0\si+b_3\st$. The class of observables defined by 
    $\A_{\rm r}$ and $\B_{\rm r}$---{\em i.e.}, $\bfX_{\rm c_2}=\{a\,\A_{\rm
    r}+b\,\B_{\rm r}\}_{a,b\in\mathbb R}$--- coincides
  with the class $\bfX_{\rm c_1}$. As a consequence, since
  $(U_{\rm C},\ket{0}\bra{0},\bfX_{\rm c_1})$ is a cloning machine,
  one has that $(U_{\rm C},\ket{0}\bra{0},\bfX_{\rm c_2})$ is a
  cloning machine too. Now, following the proof of (\ref{1pclass}), 
  one can easily show that $\forall \A$ there 
  exists a
  unitary $W_\A$ such that $\A=W_\A^\dag\A_{\rm r}W_\A$. The corresponding
  transformation on $\B_{\rm r}$ reads as follows
  \begin{eqnarray}
  W_\A^\dag{\B_{\rm r}}W_\A =b_0\si+b_3n_2\sin\theta\su
  -b_3n_1\sin\theta\sd +b_3\cos\st
  \label{BR}\;.
  \end{eqnarray}
  Together with (\ref{ABcomm}), Eq. (\ref{BR}) explicitly
shows that the observable $\B$, defined as $\B=W_\A^\dag{\B_{\rm
    r}}W_\A$, is the most general observable commuting with $\A$.
The proof is thus completed by unitary covariance and recalling that $U_\A$ is
defined as $U_\A=(W_\A^\dag \otimes W_\A^\dag) \:U_C \:(W_\A\otimes
\iid)$.
\section{Noncommuting observables and joint measurements}
\label{s:CloJM}
As we already pointed out in the previous sections there are no
cloning machines for a two--parameter class of noncommuting
observables (Theorem \ref{noclo2}).  Therefore, a question arises
on whether, analogously to state--cloning, we may introduce the
concept of approximate cloning machines, {\em i.e.} cloning of
observables involving added noise.  Indeed this can be done and
optimal approximate cloning machines corresponding to minimum
added noise may be found as well. \par An approximate cloning
machine for the class of observables $\bfX$ is defined as the
triple $(U,\vp,\bfX)_{\rm apx}$ such that $\bX_1=\bX/g_1$ and
$\bX_2=\bX/g_2$ {\em i.e.}
\begin{align}
\hbox{Tr}_1\left[ \varrho\: \X\right] 
= g_1 \hbox{Tr}_{12}\left[ R\: \X \otimes I \right] = g_2 
\hbox{Tr}_{12}\left[ R\: I \otimes \X\right]
\label{appcl}
\:,
\end{align}
for any $\X \in \bfX$.
The quantities $g_j$, $j=1,2$ are independent on the input state and 
are referred to as the noises added by 
the cloning process. 
\par
Let us begin by again considering the class $\bfX_{\rm nc}=
\{x_1\su+x_2\sd\}_{x_1,x_2\in\mathbb R}$. 
By using the decomposition of a generic $\rmSU (4)$ matrix in
\refeq{SU4} one may attempt to find an approximate cloning machine
considering only the action of the entangling kernel $U_{\rm E}$.
Unfortunately, it can be shown that no $U_{\rm E}$, $g_1$ and $g_2$
exist which realize approximate cloning for $\vp=\ket{0}\bra{0}$.
A further single-qubit transformation should be introduced after 
$U_{\rm E}$. In particular, the unitary
$F=i/\sqrt{2}(\sigma_1 + \sigma_2)$
flips the Pauli matrices  
$\su$ and $\sd$ ({\em i.e.}, $F^\dag\sigma_{1,2}\,F=\sigma_{2,1}$)
and permits the realization of an approximate cloning machine. 
Indeed, the unitary $$T=(\iid\otimes F)U_{\rm nc} \qquad
  U_{\rm nc}=e^{i\frac\theta2 (\su\otimes\su-
    \sd\otimes\sd)}\:,$$
realizes the approximate cloning machine 
$(T,\ket{0}\bra{0},\bfX_{\rm nc})_{\rm apx}$ with added noises
\begin{eqnarray}
g_1=\frac1{\cos\theta} \qquad g_2=\frac1{\sin\theta}\:.
\label{addn2}
\end{eqnarray}
In order to 
prove the cloning properties of $T$ let us start from the 
unitary $(\iid\otimes F)U_{\rm
E}$, where $U_{\rm E}$ is a generic entangling unitary of the form given 
in Eq. (\ref{SU4}). Then, by imposing approximate cloning 
for any $\X\in\bfX_{\rm nc}$, one obtains the following system of Equations:
\begin{subequations}\label{nccm_aux}
\begin{align}
  g_1\trd[(\iid\otimes\vp)\,U_E^\dag\,(\su\otimes\iid)\,U_E]&=\su  \\
  g_1\trd[(\iid\otimes\vp)\,U_E^\dag\,(\sd\otimes\iid)\,U_E]&=\sd  \\
  g_2\trd[(\iid\otimes\vp)\,U_E^\dag\,(\iid\otimes\sd)\,U_E]&=\su  \\
  g_2\trd[(\iid\otimes\vp)\,U_E^\dag\,(\iid\otimes\su)\,U_E]&=\sd
\:.\end{align}
\end{subequations}
System (\ref{nccm_aux}) admits the solution
$\theta_1=-\theta_2=\theta/2$, $\theta_3=0$---{\em i.e.}, $U_E 
\equiv U_{\rm nc}$ with $\theta$ free parameter---with 
$g_1=1/\cos\theta$ and $g_2=1/\sin\theta$. 
Notice that other solutions for the $\theta_{1,2,3}$'s
parameters may be found, which however give the same added noise as
the one considered above.
\par
Remarkably, similar cloning machines may be obtained for any class of
observables generated by a pair of operators unitarily equivalent
to $\su$ and $\sd$.
In fact, given the two-parameter classes of noncommuting observables defined as
$\bfX_{\rm V}=\{c\,\C+d\,\D\}_{c,d\in{\mathbb R}}$, with $\C=V^\dag\su
V$, $\D=V^\dag\sd V$ and $V$ generic unitary one has that
an approximate cloning machine is given by the triple 
$(U_V,\ket{0}\bra{0},\bfX_{\rm V})_{\rm apx}$, with $U_V=(V^\dag
\otimes V^\dag) (\iid\otimes F)\:U_{\rm nc} \:(V\otimes \iid)$, 
with added noises $g_1=1/\cos\theta$ and $g_2=1/\sin\theta$. 
The statement easily follows from the fact that 
$(T,\ket{0}\bra{0},\bfX_{\rm nc})_{\rm apx}$ is a cloning
machine and from unitary covariance. 
Similar results hold for any class of observables 
unitarily generated by any 
pair of (noncommuting) Pauli operators. 
\par
A question arises about optimality of approximate cloning machines
for noncommuting observables. In order to assess the quality and to 
define optimality of a triple $(U,\vp,\bfX)_{\rm apx}$ we consider 
it as a tool to perform a joint measurements of noncommuting 
qubit observables \cite{tri}. For example, consider the cloning machine   
 $ (T,\ket{0}\bra{0},\bfX_{\rm nc})_{\rm apx}$
 and suppose to measure $\su$ and $\sd$ on the two qubits at the output. 
We emphasize that the cloning machine 
$ (T,\ket{0}\bra{0},\bfX_{\rm nc})_{\rm apx}$
clones every observable belonging to $\bfX_{\rm nc}$, while we are now 
considering only the observables $\su$ and $\sd$ which, in a sense, 
generate the class. We have that the measured expectation values 
of $\su$ and $\sd$ at the output are given by 
$\avg{\sigma_h}_{\rm m}=g_h\avg{\sigma_h}$ (with $h=1,2$),  
where the $\avg{\sigma_h}$'s are the input mean values. 
It follows that the measured uncertainties ($\Delta O=\avg{O^2}-\avg{O}^2$) 
at the output are given by $$\Delta_{\rm m}\sigma_h=g_h^2\Delta_{\rm
i}\sigma_h$$ 
where $\Delta_{\rm i}\sigma_h$ denote the intrinsic uncertainties for 
the two quantities at the input. Since for any Pauli operators 
we have $\sigma_h^2=\iid$ one may
rewrites
\begin{align}
\Delta_{\rm m}\su &=\tan^2\theta+\Delta_{\rm i}\su \\ 
\Delta_{\rm m}\sd &=\cot^2\theta+\Delta_{\rm i}\sd\:. 
\end{align}  
As a consequence, the measured uncertainty product is given by:
  $$ \Delta_{\rm m}\su\Delta_{\rm m}\sd=\Delta_{\rm i}\su\Delta_{\rm i}\sd
+\cot^2\theta\Delta_{\rm i}\su+\tan^2\theta\Delta_{\rm i}\sd+1\:.$$
Since the arithmetic mean is bounded from below by the
geometric mean we have
  $$\cot^2\theta\Delta_{\rm i}\su+\tan^2\theta\Delta_{\rm i}\sd 
  \ge 2\sqrt{\Delta_{\rm i}\su\Delta_{\rm i}\sd}\:,$$
  with the equal sign iff $\Delta_{\rm i}\su=\tan^4\theta\Delta_{\rm
  i}\sd$, then it follows that
  $$\Delta_{\rm m}\su\Delta_{\rm m}\sd\ge
  \left(\sqrt{\Delta_{\rm i}\su\Delta_{\rm i}\sd}+1\right)^2\:.$$
  If the initial signal is a minimum uncertainty state---{\em i.e.}, 
  $\Delta_{\rm i}\su\Delta_{\rm i}\sd=1$---one finally has that the 
  measured uncertainty product is bounded by 
  $\Delta_{\rm m}\su\Delta_{\rm m}\sd\ge4$.
  Notice that 
  an optimal joint measurement corresponds to have
  $\Delta_{\rm m}\su\Delta_{\rm m}\sd = 4$. In our case this 
  is realized when $\theta$ is chosen such that
  $\tan^4\theta=\Delta_{\rm i}\su/\Delta_{\rm i}\sd$.
Therefore, since 
 $ (T,\ket{0}\bra{0},\bfX_{\rm nc})_{\rm apx}$
adds the minimum amount of noise in a joint measurement 
performed on minimum uncertainty states we conclude that it is 
an optimal approximate cloning machine for the class under
investigation. An optimal approximate cloning machine for the more
general class $\bfX_{\rm gnc}$ may be also defined, using the 
concept of joint measurement for noncanonical observables \cite{tri}.
\par
Let us now consider the comparison with a joint measurement of
$\su$ and $\sd$ performed with the aid of an optimal universal cloning
machine for states \cite{buz}. 
It is easy to show that the best result in this
case is given by $\Delta_{\rm m}\su\Delta_{\rm m}\sd=\frac92$,
indicating that cloning of observables is more effective than cloning
of states to perform joint measurements (for the case of three
observables see Refs. \cite{hil,pva}). In fact, a symmetric cloning 
machine for states shrinks the whole Bloch vector $\boldsymbol s$ by 
a factor $\frac23$, whereas a cloning machine for observables shrinks 
the components $s_1$ and $s_2$ of $\boldsymbol s$ only by a factor 
$1/\sqrt2$ (considering equal noise $g_1=g_2=\sqrt2$). Notice that 
such a behavior is different from what happens in the case of 
continuous variables, for which the optimal covariant cloning of coherent 
states also provides the optimal joint measurements of two conjugated 
quadratures \cite{cer}. This is due to the fact that coherent states are 
fully characterized by their complex amplitude, that is  by the 
expectation values of two operators only, whereas 
the state of a qubit requires the knowledge of the three
components of the Bloch vector.
\par 
As a final remark we notice that if the requirement of
universality is dropped, then cloning machines for states
can be found that realize optimal approximate cloning 
of observables. For example, an approximate 
cloning machine for the  two--parameter class $\bfX_{\rm nc}$
can be obtained by considering a phase--covariant cloning for
states \cite{phc}. In order to clarify this point, let us recall 
that a phase--covariant cloning machine for states of the form 
uses a probe in the $|0\rangle$ state and performs the following
transformation:
\begin{align}
 |0\rangle |0\rangle & \rightarrow  |0\rangle|0\rangle
 \nonumber \\
 |1\rangle |0\rangle & \rightarrow  
  \cos\theta|1\rangle|0\rangle+
  \sin\theta|0\rangle|1\rangle)\,,\label{eqc}
 \end{align}
where, in general, $\theta\in[0,2\pi]$. 
If we now consider the $\bfX_{\rm nc}$ class it is straightforward to
show that Eqs. (\ref{appcl}) are satisfied for any
$\X\in\bfX_{\rm nc}$ using the machine (\ref{eqc}), 
with the optimal added noises given by Eq. (\ref{addn2}).
This can be intuitively understood by considering that a phase covariant
cloning machine extracts the optimal information about states lying on
the equatorial plane of the Bloch sphere, which in turn include the
eigenstates of $\sigma_1$ and $\sigma_2$. 
\section{Conclusions}\label{s:outro}
We addressed the encoding of information in the statistics of a
set of observables, independently on the quantum state at the
input. In a network of this kind there is no need of a precise
control of the coding stage whereas, on the other hand, each gate
should be {\em transparent}, {\em i.e} should preserve the
statistics of the transmitted observables. To this aim the
concepts of exact and approximate cloning for a class of
observables have been introduced and developed. Explicit
realizations of cloning machines have been found for classes of
commuting observables, which in turn realize quantum
non-demolition measurements of each observable belonging to the
class.  Two no-cloning theorems for observables have been derived
which subsumes both the no--cloning for states and the
impossibility of joint measurement of noncommuting observables
without added noise.  In addition, approximate cloning machines
for classes of noncommuting observables have been also found,
which realize optimal joint measurements.  We found that cloning
of observables is more effective than universal cloning of states
to perform joint measurements of $\su$ and $\sd$ since a
symmetric cloning machine for states shrinks the whole Bloch
vector  whereas a cloning machine for observables shrinks only
two components by a smaller factor.  On the other hand, if one
restricts attention to non--universal cloning of states, then
approximate cloning of observables is equivalent to
phase-covariant cloning of states.
\section*{Acknowledgments}
This work has been supported by MIUR through the project PRIN-2005024254-002.


\begin{thebibliography}{99}
\bibitem{nt1} The roles of $a_{1,2,3}$ can be interchanged to
  avoid singularities.
\end{thebibliography}
\end{document}